\documentclass[journal=jacsat,manuscript=article]{achemso}
\setkeys{acs}{articletitle = true}

\graphicspath{ {./images/} }

\usepackage[version=3]{mhchem} 

\usepackage{multicol}
\usepackage{graphicx} 
\usepackage{changepage}
\usepackage{amsmath}
\usepackage{verbatim}
\usepackage{color}

\usepackage{soul}

\author{Ioannis Karageorgiou}
\affiliation{Yusuf Hamied Department of Chemistry, University of Cambridge, CB2 1EW Cambridge, UK}

\author{Angelos Michaelides}
\affiliation{Yusuf Hamied Department of Chemistry, University of Cambridge, CB2 1EW Cambridge, UK}
\email{am452@cam.ac.uk}

\author{Fabian Berger}
\affiliation{Yusuf Hamied Department of Chemistry, University of Cambridge, CB2 1EW Cambridge, UK}
\email{fb593@cam.ac.uk}

\title[\texttt{achemso}]
{Mechanisms for the Formation of Active Sites in Single-Atom Alloys}

\begin{document}

\clearpage

\begin{abstract}

Reactive dopant atoms embedded in inert host metal surfaces define the active sites in single-atom alloys (SAAs), yet SAA synthesis remains challenging.
To address this, we elucidate how dopant adatoms deposited on Cu and Ag surfaces become incorporated into the metal and identify periodic trends from early to late transition metals (TMs) using density functional theory.
Adatoms diffuse nearly freely across terraces, as diffusion barriers are small, whereas direct incorporation into terraces is unfavourable.
In line with conventional wisdom, step edges and kink sites strongly facilitate dopant incorporation, confirming their critical role in alloy formation.
Attachment of adatoms to steps and kinks from the lower terrace is favoured.
Incorporation then proceeds either from this attached state or when adatoms approach a step edge from above, where reactions often proceed without barrier.
Incorporation barriers are generally lower for early and central TMs, increase towards late TMs, and are slightly higher on Cu than on Ag surfaces.
Repulsive interactions between Pd adatoms and dopants explain the experimental observation that a dopant-rich brim on the upper terrace of Cu surfaces inhibits incorporation from above.
In contrast, attractive interactions, as found for Ru, anchor diffusing adatoms (even on terraces) and promote the formation of adatom islands, yet hinder incorporation next to the dopant and may impede the growth of embedded dopant clusters.
By rationalising periodic trends and experimental observations, we show how specific surface sites and adatom--dopant interactions shape dopant incorporation, offering guidance on the surface environments most conducive to SAA synthesis for different dopant elements.

\end{abstract}

\clearpage

\section{Introduction}

Heterogeneous catalysis drives most industrial chemical processes,\cite{demand_catalysis,ma2006heterogeneous} but conventional solid catalysts are often constrained by linear scaling relations, resulting in a trade-off between activity and selectivity. \cite{norskov2002universality,michaelides2003identification,sabatier1920catalyse,che_nobel_2013}
Single-atom alloys (SAAs), in which atomically dispersed reactive dopants are embedded in relatively inert host metals, can overcome this limitation.\cite{hannagan2020single,kyriakou_isolated_2012,reocreux2021one,cui2018bridging,yang_single-atom_2013,lee2022dilute}
This architecture spatially separates active reaction sites from inert desorption sites, allowing formed products to spill over, introducing bifunctionality. \cite{tierney_hydrogen_2009,reocreux2021one}
Combined with the unique free-atom-like electronic structure of the dopant sites, SAAs can enhance activity while maintaining high selectivity.\cite{greiner2018free,thirumalai_investigating_2018,rosen2023free,berger_when_2025,spivey_selective_2021}
With tunable composition and low dopant concentrations, typically only a few percent in the surface layer, SAAs also maximize the efficiency of costly metals, making them a highly promising class of solid catalysts.

As SAAs are exciting catalysts, most experimental and computational studies focus on their catalytic performance, using samples and models with already formed active sites consisting of embedded dopants. \cite{hannagan2020single, chen_electronic_2022,huang2024origin, zhang2021single, schumann_ten-electron_2024, berger_bringing_2024, berger_when_2025, reocreux2021one, reocreux2022stick, schumann2021periodic}
This has been demonstrated with great success for selective hydrogenation,\cite{kyriakou_isolated_2012,lucci_selective_2015} selective oxidation,\cite{jalil_nickel_2025} dehydrogenation,\cite{hannagan_first-principles_2021} and CO oxidation.\cite{therrien2018atomic}
Much less attention, however, has been devoted to understanding how these active dopant sites form in the first place.\cite{wang_surface_2020, troling_toward_2017, lim_automated_2019}
Elucidating this process is not only of fundamental mechanistic interest but could also inform improved synthesis strategies, which remain a major bottleneck due to experimental challenges.\cite{zhang_recent_2024, jia2024challenges, asikin2021single}

Among the methods available for synthesising highly dilute metal alloys,\cite{hannagan2020single,boucher2013single, giannakakis2018niau, wong2017synthesis, hakim2015synthesis} physical vapor deposition (PVD) is widely employed.\cite{baptista2018sputtering,shi_formation_2021,tierney_atomic-scale_2009,lucci_atomic_2014}
In this work, we model conditions representative of PVD.
The synthesis of SAAs typically begins with the deposition of a trace amount of dopant onto the host metal surface, initially forming adatoms that reside on the surface before becoming incorporated into the alloy.
From this initial state, several processes can occur, which we elucidate and relate to experimental and computational studies.\cite{bach_aaen_submonolayer_1998,lucci_atomic_2014,bellisario_importance_2009,mo_electronic_2008, kim_transition-pathway_2007,troling_toward_2017}
Using density functional theory (DFT) with the optB86b-vdW exchange–correlation functional, \cite{optB86b-vdW} we investigate the series of 4\textit{d} transition metal (TM) dopants to identify periodic trends and include Cu and Ag hosts to assess the influence of the host element.
In addition to the ideal and commonly studied terrace (111) facet, we model step edges using (211) and (322) surfaces and construct additional models for kink defects derived from these facets.
These surfaces correspond to one of the two step-edge types found on a (111) surface and are the type most widely investigated.\cite{li_potential_1994, kim_transition-pathway_2007,trushin_step_1997, lim_automated_2019, villarba_diffusion_1994, ali_mechanism_2022, feibelman1998interlayer}

\section{Results and Discussion}
Mechanisms for the diffusion, attachment, and incorporation of adatoms into metal surfaces have been investigated previously,\cite{wang_surface_2020,hayat_diffusion_2010,kim_transition-pathway_2007,feibelman_diffusion_1990,troling_toward_2017,wrigley_surface_1980,lim_automated_2019,bulou_mechanisms_2005,halim_multi-scale_2021,stumpf1994theory,lim_evolution_2020,kim_transition-pathway_2007-1,li_potential_1994,mo_electronic_2008,bellisario_importance_2009,villarba_diffusion_1994,li_predicted_1996,ali_mechanism_2022,henkelman_multiple_2003,trushin_step_1997} but typically only for very few, specific combinations of adatom and host elements and often focusing on a limited subset of possible pathways.
As a result, a broader, unified picture of how adatoms evolve into embedded dopants and how these mechanisms vary across the periodic table has remained incomplete.
By investigating relevant pathways using a consistent computational framework, we establish this coherent mechanistic understanding and identify clear periodic trends.
Full computational details are provided in Section S1 of the Supporting Information.

\begin{figure*}[htp]
\centering
  \includegraphics[width=\textwidth]{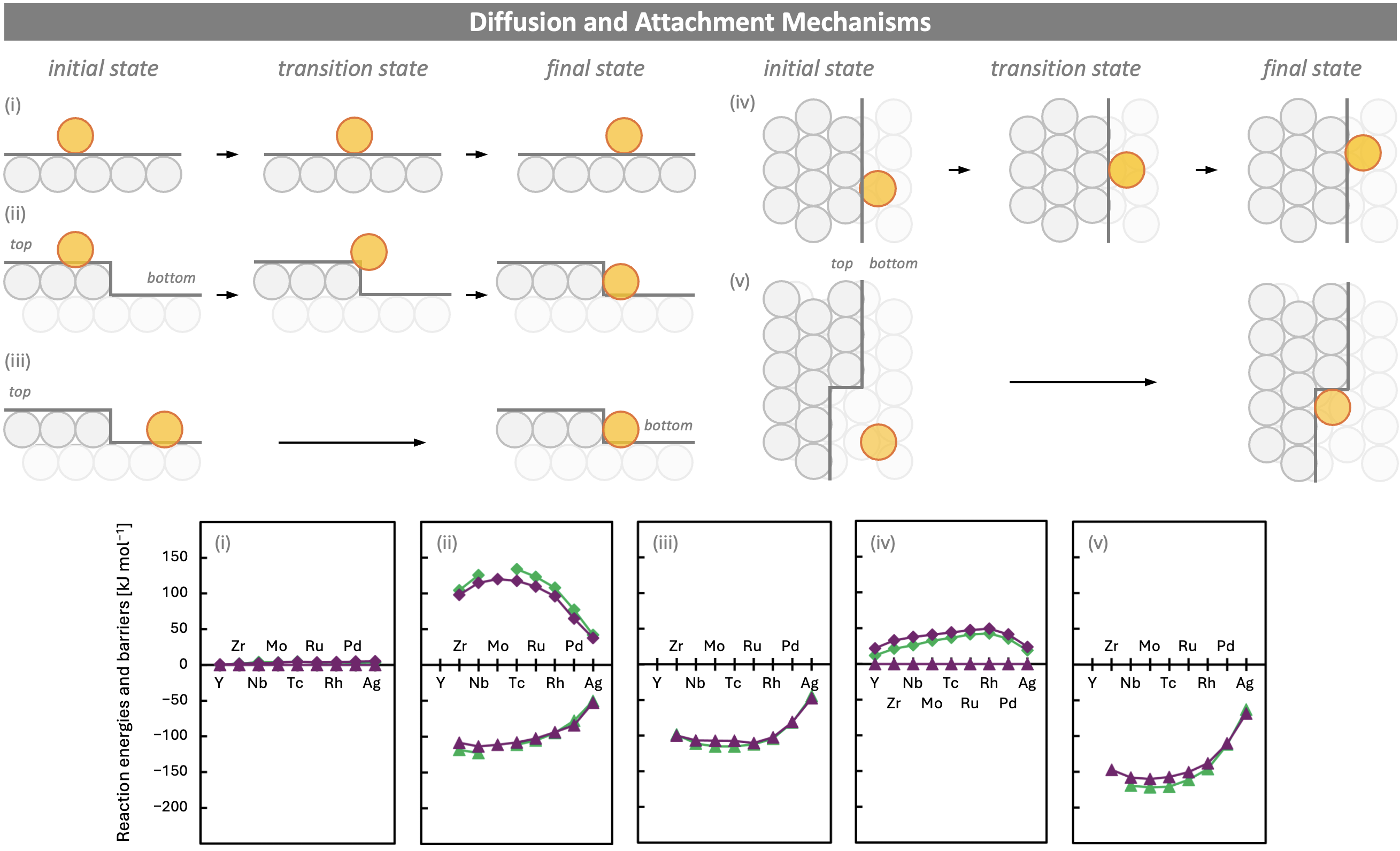}
  \caption{
    Schematic overview of the diffusion and attachment mechanisms considered in this work, together with their periodic trends in reaction energies (triangles) and reaction barriers (diamonds) for 4\textit{d} TM adatoms on Cu (green) and Ag (purple) surfaces.
    The left configuration corresponds to the initial reactant, followed by the transition state (if present) and the final product.
    Illustrated mechanisms are:
    (i) diffusion of an adatom across a terrace;
    (ii) hopping of an adatom from the upper to the lower terrace of a step edge;
    (iii) attachment of an adatom to a step edge from the lower terrace;
    (iv) diffusion of an adatom already attached to a step edge along the edge; and
    (v) attachment of an adatom to a kink from the lower terrace.
    Adatoms are shown in orange and host atoms in grey; atoms in the lower terrace appear in light grey and those in the upper terrace in grey.
    The surface, step edge, and kink are indicated with a dark grey line.
    Surface dimensions are for illustrative purposes only; simulation cell sizes are provided in Section S1 of the Supporting Information.
    Connecting lines are for illustrative purposes only.
  }
  \label{fgr:diffusion_attachment}
\end{figure*}

An adatom deposited on a surface can diffuse across the terrace, attach to defects such as step edges and kink sites, and eventually become incorporated into the surface. 
We first discuss the diffusion and attachment processes shown in Figure~\ref{fgr:diffusion_attachment}, as these determine how adatoms move across the surface and reach defects. 
We then investigate the different incorporation mechanisms on terraces, step edges, and kinks and identify the trends that govern their energetics. 
Finally, we combine these insights to determine which reaction pathways dominate and illustrate how already embedded dopants can influence adatom diffusion and incorporation.

\textbf{Adatom diffusion and attachment to step edges and kinks.}
Starting from an adatom deposited on a terrace site, adatom diffusion across terraces,\cite{minkowski2015diffusion, troling_toward_2017,hayat_diffusion_2010,kim_transition-pathway_2007, feibelman_diffusion_1990} as illustrated in mechanism (i), is extremely fast.
The energy differences between adsorption sites and the corresponding diffusion barriers (diamonds) are predicted to be below 5~kJ~mol$^{-1}$.
Such small barriers enable adatoms to explore terrace regions quickly.
Relative energies and barriers are provided in Section~S2 of the Supporting Information.

Step edges separate an upper and a lower terrace and can therefore be approached by adatoms from either side.
The descending side connects the upper to the lower terrace, while the ascending side connects the lower to the upper terrace.
Kink sites correspond to local irregularities in a step, typically formed by host atoms protruding from the edge.
These structural features play an important role in guiding adatom motion and determining how adatoms attach to or incorporate into the surface.

When an adatom approaches a step edge from the upper terrace, it may hop down onto the lower terrace and attach to the ascending step,\cite{li_potential_1994, mo_electronic_2008,bellisario_importance_2009,stumpf1994theory} as shown in mechanism (ii).
This process exhibits comparatively large barriers with an inverse U-shaped trend.
Because the adatom loses coordination in the transition state, elements that form stronger bonds with the host show higher barriers, in some cases far exceeding 100~kJ~mol$^{-1}$.
Very late TMs, which form weaker bonds with the host,\cite{berger_when_2025} exhibit smaller barriers, as low as about 40~kJ~mol$^{-1}$ for Ag adatoms, which may be surmountable at moderate temperatures.

Adatoms approaching a step edge from the lower terrace readily attach to the ascending step.
This process is shown in mechanism (iii).
The barriers for this attachment do not exceed those for terrace diffusion and can become barrierless once the adatom is close to the step.
Since attachment increases the coordination of the adatom, it is strongly exothermic, releasing up to 115~kJ~mol$^{-1}$ for Mo.
The exothermicity decreases along the dopant series (triangles), reaching 44~kJ~mol$^{-1}$ for Ag on Cu, consistent with the weaker bonds formed by late TMs.

Once an adatom is attached to a step edge on the lower terrace, it may diffuse along the edge,\cite{halim_multi-scale_2021, stumpf1994theory} as illustrated in mechanism (iv).
The barriers for this process are much higher than those for terrace diffusion, reaching up to about 50~kJ~mol$^{-1}$ for Rh on Ag.
As a result, diffusion along step edges is slow.
At elevated temperatures relevant for catalysis, these barriers may be overcome, allowing adatoms to migrate along the step until they encounter a kink site.
The barriers are smaller for late TMs.
For the coinage host metal adatoms, step-edge diffusion barriers are only about 20~kJ~mol$^{-1}$, which provides a possible mechanism for the growth of step edges.

When an adatom approaches a kink from the lower terrace,\cite{chen_electronic_2022, ali_mechanism_2022} it attaches strongly, as shown in mechanism (v).
Similar to attachment at a step edge, the barriers for this process do not exceed those for terrace diffusion and can become barrierless once the adatom is close to the kink.
Because kink sites offer the largest number of neighbouring host atoms for an adatom, attachment is highly exothermic, releasing up to 172~kJ~mol$^{-1}$ for Mo on Cu.
The stabilisation decreases along the dopant series as the strength of the bonds formed between the adatom and the host atoms reduces, and eventually reaches 63~kJ~mol$^{-1}$ for Ag on Cu.
The stronger binding at kink sites compared with straight step edges reflects the increased coordination.
Overall, adatoms preferentially occupy sites that maximise the number of neighbouring host atoms: from terraces, where they interact only with atoms beneath, to step edges, where they gain additional neighbours on one side, and finally to kinks, where they interact with host atoms from two sides.
This preference is more pronounced for elements that form stronger bonds with the host.

\begin{figure*}[htp]
\centering
  \includegraphics[width=\textwidth]{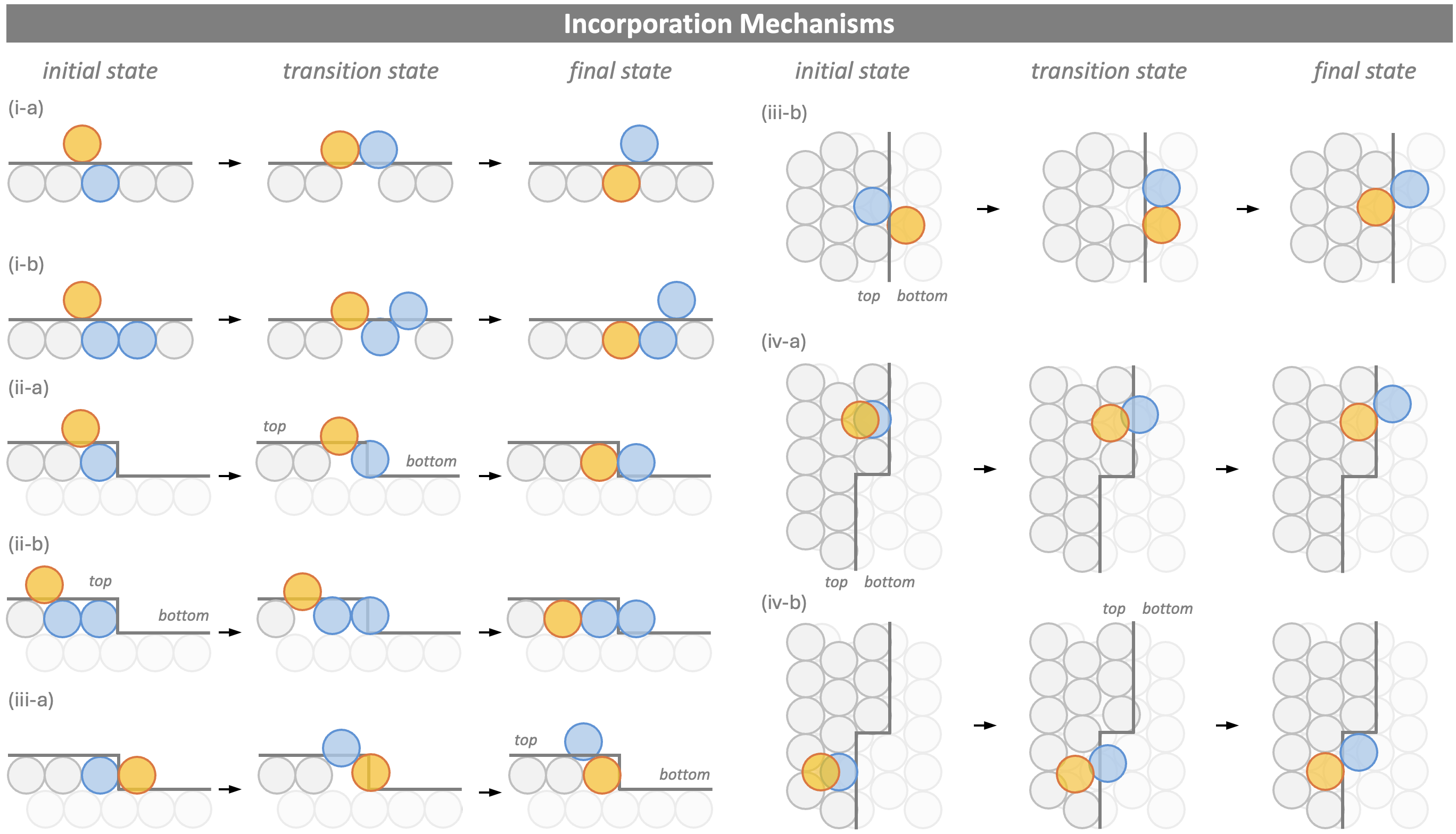}
  \caption{
    Schematic illustration of the different mechanisms for the incorporation of an adatom as an embedded dopant considered in this work.
    The left configuration corresponds to the initial reactant, followed by the transition state structure and the final product.
    The illustrated mechanisms are:
    (i-a) direct incorporation into a terrace site involving the adatom and one host atom;
    (i-b) incorporation into a terrace site involving the adatom and a concerted motion of two host atoms;
    (ii-a) incorporation of an adatom approaching the step edge from above;
    (ii-b) incorporation of an adatom into the neighbouring row of the step edge, involving the concerted motion of two host atoms;
    (iii-a) incorporation of an adatom attached to the step edge on the lower terrace, forming a host-atom adatom attached to the upper terrace of the edge;
    (iii-b) incorporation of an adatom attached to the step edge on the lower terrace, forming a host-atom kink attached to the edge from below;
    (iv-a) incorporation of an adatom approaching the step edge from above near a kink;
    (iv-b) incorporation of an adatom approaching the step edge from above at a kink.
    The adatom and dopant are shown in orange, the moving host metal atoms in blue, and the adjacent host metal atoms in grey.
    Host atoms in the lower terrace are shown in light grey, while those in the upper terrace are shown in grey.
    The surface, step edge, and kink are indicated with a dark grey line.
    Surface dimensions are for illustrative purposes only; the actual simulation cell sizes are provided in Section~S1 of the Supporting Information.
  }
  \label{fgr:incorporation_mechanisms}
\end{figure*}

\textbf{Adatom incorporation mechanisms.}
Once an adatom has diffused across the surface and interacted with step edges or kink sites, it may become incorporated into the lattice.
The incorporation mechanisms considered in this work are illustrated in Figure~\ref{fgr:incorporation_mechanisms} and occur on terraces, step edges, and kinks.
These processes embed the adatom into the surface and expel a host atom, which becomes a new adatom.
Depending on where this expelled atom is located, it may remain attached to the ascending step edge or kink, or it may start diffusing or undergo a consecutive incorporation when expelled onto the upper terrace.

Once an adatom is on the surface, incorporation can proceed through several pathways.
On terraces, the adatom (orange) may directly replace a host atom in the surface layer,\cite{troling_toward_2017, lim_automated_2019,wrigley_surface_1980} expelling it as a new adatom, following mechanism (i-a).
Alternatively, incorporation can proceed through a concerted motion of the adatom and two host atoms (blue),\cite{lim_automated_2019, bulou_mechanisms_2005} shown in mechanism (i-b), which reduces the steric constraints associated with direct exchange.
In both cases, the adatom becomes embedded in the terrace, while the expelled host atom remains on the surface: in the former case directly attached to the newly incorporated dopant, and in the latter bonded to neighbouring host atoms.

When an adatom approaches a step edge from the upper terrace, incorporation can occur directly at the edge.
In mechanism (ii-a), the adatom replaces a host atom in the step edge,\cite{villarba_diffusion_1994,li_potential_1994,mo_electronic_2008, bellisario_importance_2009, li_predicted_1996,ali_mechanism_2022,lim_automated_2019,stumpf1994theory} which is expelled onto the lower terrace as a new adatom.
In mechanism (ii-b), incorporation proceeds through a concerted motion of the adatom and two host atoms,\cite{henkelman_multiple_2003} placing the dopant in the neighbouring row next to the edge and again expelling a host atom onto the lower terrace.
Attempts to localise a concerted mechanism involving the adatom and three host atoms were often unsuccessful; instead, the reaction path breaks into two steps, with the terrace concerted exchange (i-b) followed by incorporation next to the edge via (ii-b).
Both identified incorporation pathways embed the dopant in the upper terrace and create a new kink site, which consists of the expelled host atom attached to the edge from below.

Adatoms that are attached to a step edge on the lower terrace can incorporate into the edge through two pathways.
In mechanism (iii-a), the attached adatom replaces a host atom in the step edge and expels it onto the upper terrace,\cite{lim_automated_2019, lim_evolution_2020, kim_transition-pathway_2007-1} where it becomes a new adatom.
Following mechanism (iii-b), the adatom again replaces a host atom, but the expelled atom remains on the lower terrace,\cite{halim_multi-scale_2021} where it forms a new kink site.
In both cases, the dopant ends up embedded in the step edge, while the expelled host atom either becomes a mobile adatom on the upper terrace or forms a kink protruding from the step edge on the lower terrace.

Approach to a kink site from the upper terrace can result in incorporation through two closely related pathways, depending on the precise location of the adatom relative to the kink.\cite{kim_transition-pathway_2007-1, trushin_step_1997, ali_mechanism_2022, li_predicted_1996}
In mechanism (iv-a), an adatom positioned near the kink replaces a host atom in the edge and expels it onto the lower terrace, producing a new kink that can initiate the growth of an entirely new row.
Following mechanism (iv-b), incorporation occurs directly at the kink, with the expelled host atom again moving to the lower terrace, but this time the kink shifts by one lattice site along the step, extending the advancing edge.
Both pathways embed the dopant in the upper terrace, yet the positioning of the expelled host atom leads to different structural outcomes. 
Mechanism (iv-a) creates a new kink, whereas mechanism (iv-b) propagates the existing step.

\begin{figure*}[htp]
\centering
  \includegraphics[width=14cm]{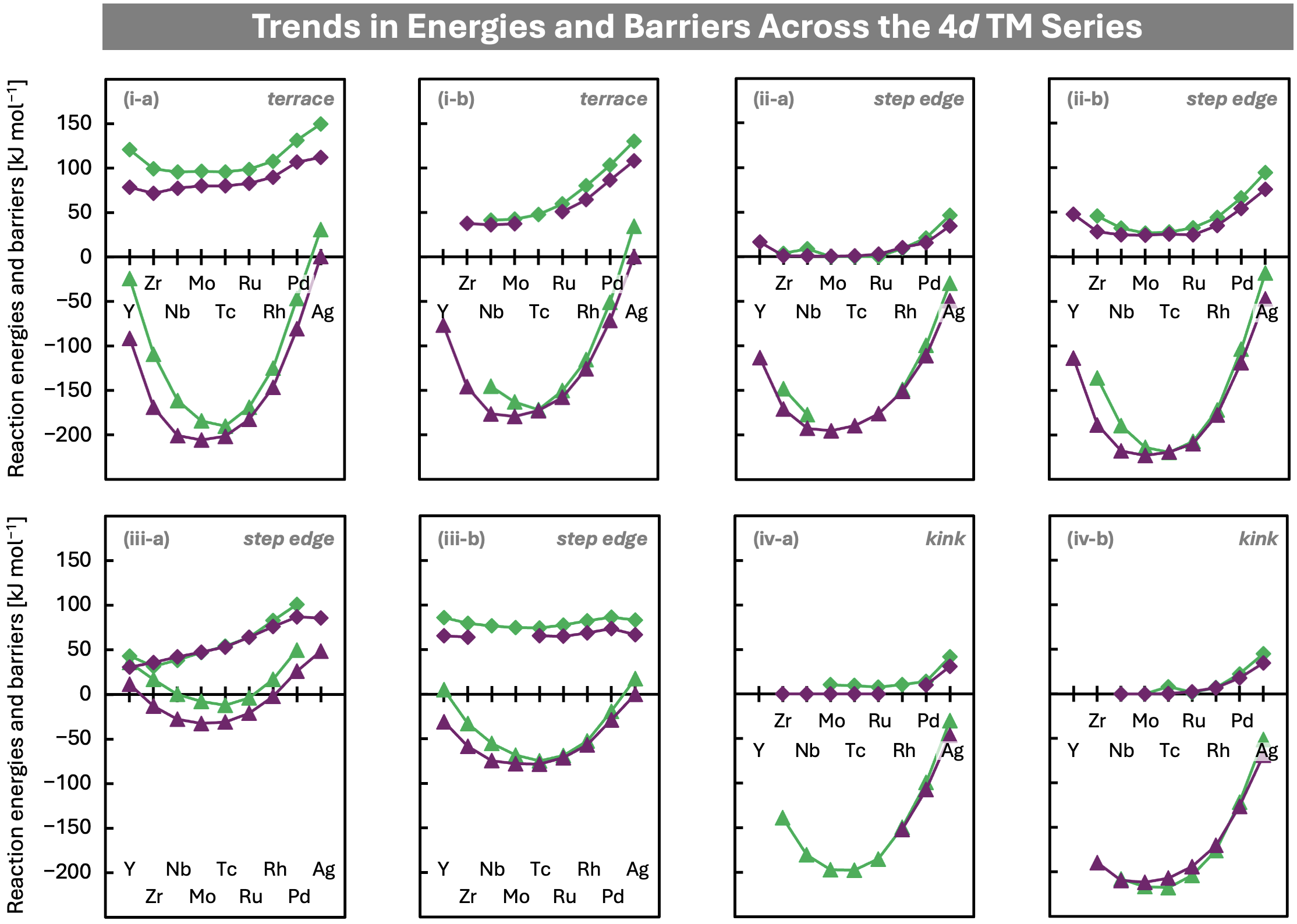}
  \caption{
    Periodic trends in the reaction energies (triangles) and reaction barriers (diamonds) for incorporation of 4\textit{d} transition metal dopants on Cu (green) and Ag (purple) host metal surfaces.
    All energies are given in kJ~mol$^{-1}$.
    While reaction energies exhibit U-shaped periodic trends typical of many SAA properties, reaction barriers display attenuated U-shaped, gradually increasing, or no significant trends across the dopant series, as discussed in the text.
    Connecting lines are provided for illustrative purposes only.
  }
  \label{fgr:periodic_trends}
\end{figure*}

\textbf{Periodic trends in the incorporation mechanisms.}
With the mechanistic picture in place, we next consider the energetics of the pathways.
Figure~\ref{fgr:periodic_trends} summarises the reaction energies (triangles) and barriers (diamonds) associated with the incorporation mechanisms discussed above, revealing clear periodic trends across the 4\textit{d} TM series.
These trends arise from variations in atomic size and bonding strength along the series: central TMs are similar in size to the host but considerably more reactive, gaining substantial stabilisation upon incorporation; early TMs are larger, introducing a size mismatch that reduces the stabilisation gained; and late TMs form weaker bonds with host atoms,\cite{berger_when_2025} and therefore exhibit less exothermic reactions and typically higher incorporation barriers.

The key factor underlying the periodic trends in reaction energies is the change in coordination experienced by the adatom during incorporation.
Among all surface sites, an atom embedded in a terrace has the largest number of direct neighbours, followed by incorporation into a step edge or kink, then attachment to a kink, attachment to a straight step edge, and finally adsorption on a terrace.
Elements that form stronger bonds with the host are therefore stabilised more strongly when their coordination increases, giving rise to the characteristic U-shaped trends in reaction energies, with central TMs exhibiting the largest stabilisation upon incorporation.
For early TMs, the energy gained by forming additional bonds is (partly) offset by steric strain arising from their larger atomic size compared with the more compact coinage metal lattice.
Late TMs, although similar in size to the host, form weaker bonds with host atoms,\cite{berger_when_2025} so increasing the number of neighbours yields only limited stabilisation.
Moreover, incorporation necessarily moves a host atom from a highly coordinated lattice position to a less coordinated surface site, and the resulting loss of coordination reduces the overall exothermicity of the reaction.
As long as the adatom forms stronger bonds with the host than the host does with itself, incorporation remains exothermic.
However, when the adatom forms weaker bonds, for example Ag incorporated into Cu, the reaction can become endothermic.

In contrast to the pronounced U-shaped reaction energies, the reaction barriers associated with the different incorporation pathways exhibit more diverse trends across the 4\textit{d} series.
Overall, three patterns emerge: (1) an attenuated U-shape, (2) a gradual increase from early to late TMs, and (3) no discernible dependence on the dopant element.
U-shaped trends arise when the moving host atom remains in close contact with the replacing adatom in the transition state (TS), as in mechanisms (i-a), (ii-a), and (ii-b) illustrated in Figure~\ref{fgr:incorporation_mechanisms}.
Late TMs, being more inert, are less able to stabilise the detaching host atom in the TS, which leads to higher barriers, whereas central TMs stabilise it more effectively through stronger bonding.
For very early TMs, the large atomic size can slightly destabilise the TS when the adatom is in close proximity and cannot relax efficiently within the confined structure of the TS.
Barriers that increase gradually from early to late TMs occur in mechanisms (i-b) and (iii-a), which may be explained by a better stabilisation of the detaching host atom in the TS by the larger early TMs when the incorporating and detaching atoms are not as close.
The weakening of bond strengths along the series then leads to increasing barriers for late TMs.
Finally, no clear periodic trends are observed when the TS involves significant detachment of the host atom before any new bonds can form with the adatom, as in mechanism (iii-b), or when the incorporation pathways are nearly barrierless for most elements, such as incorporation from above into step edges or kinks following mechanisms (ii-a), (iv-a), and (iv-b).
These pathways show only a slight increase in barriers for the latest TMs, consistent with the fact that these elements form the weakest bonds.

These trends are consistent across both host metals.
Barriers on Cu surfaces are slightly higher, reflecting the stronger Cu--Cu bonds that need to be broken during the detachment of a host atom accompanying adatom incorporation.
Reactions are also slightly less exothermic for Cu, which can be attributed to its shorter interatomic distances compared with Ag.
Because the incorporated dopants are 4\textit{d} TMs, which are larger than the 3\textit{d} Cu atoms but more similar in size to Ag, the size mismatch in Cu introduces additional strain, thereby reducing the energetic stabilisation gained upon incorporation.

Taken together, these periodic trends in reaction energies and barriers provide a detailed understanding of how adatom properties influence individual incorporation mechanisms, but they do not, on their own, determine how dopants actually enter the surface during SAA formation.
To resolve which pathways operate in practice, it is also necessary to consider how frequently adatoms reach the relevant reactant states in addition to the intrinsic reaction barriers.

\textbf{Dominant incorporation at step edges and kinks.}
Assuming an adatom is deposited on a terrace site, which is the most common surface type, it will first diffuse rapidly, since terrace diffusion is much faster than any direct incorporation process into the terrace.
As a result, adatoms typically explore large areas of the surface and encounter defect sites before terrace incorporation can occur.
At elevated temperatures, however, the concerted terrace incorporation pathway (i-b) may become accessible for early TMs, as the barriers can decrease to about 36~kJ~mol$^{-1}$ for Nb on an Ag surface.
This indicates that concerted mechanisms deserve more attention in mechanistic studies than they have received so far. \cite{bulou_mechanisms_2005,lai_automatic_2025}

\begin{figure*}[ht]
\centering
  \includegraphics[width=\textwidth]{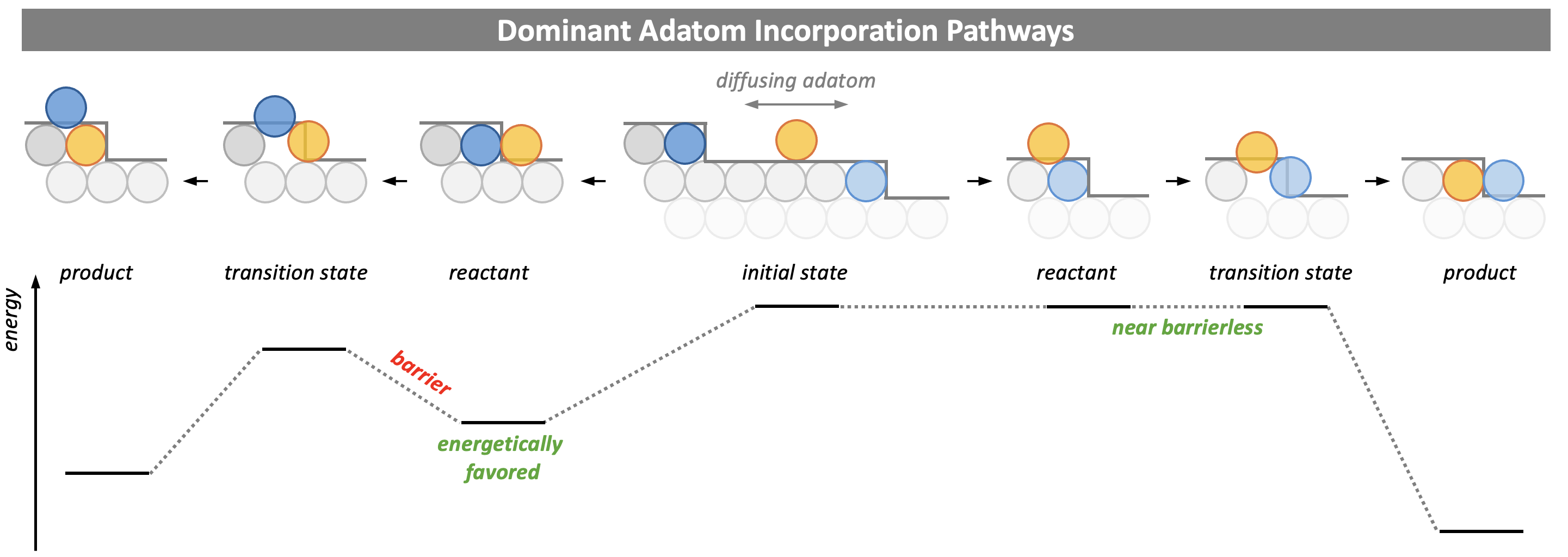}
  \caption{
    Schematic illustration of the two most dominant pathways for the incorporation of adatoms as dopants.
    Initial deposition of an adatom on a terrace leads to rapid diffusion with small barriers.
    Assuming a random walk of the adatom, approaches to a defect such as a step edge from above or from below are equally likely.
    If approaching a step edge from above, incorporation into the step edge is barrierless for most dopant elements and leads to the formation of a host atom kink attached to the edge on the lower terrace, according to mechanism (vi-a).
    If approaching a step edge from below, the adatom attaches and becomes trapped at the edge until it incorporates following mechanism (iv-b), as the energy required for detachment is typically higher than for incorporation.
    Schematic energy profiles are shown beneath the structural representations.
    The adatom and dopant are shown in orange, the moving host metal atoms in blue, and the adjacent host metal atoms in grey.
    Host atoms in the lowest terrace are shown in light grey, those in the central terrace in grey, and those in the uppermost terrace in dark grey.
    The surface, step edge, and kink are indicated with dark grey lines.
    Surface dimensions are for illustrative purposes only; the actual simulation cell sizes are provided in Section~S1 of the Supporting Information.
  }
  \label{fgr:dominant_pathways}
\end{figure*}

All step edges have an ascending and a descending side.
Hence, diffusing adatoms have equal probability to approach an edge from the lower or the upper terrace.\cite{bellisario_importance_2009}
Each of these approaches can initiate one of the two dominant pathways illustrated in Figure~\ref{fgr:dominant_pathways}.
Thermodynamically, attachment from below is strongly favoured, and attachment to a kink is even more stabilising.

The barriers for diffusion along the edge, following mechanism (iv), are below 50~kJ~mol$^{-1}$, with the highest value found for the late TM Rh on Ag.
For early TMs, these barriers can be as low as 12~kJ~mol$^{-1}$, for example for Y on Cu.
This diffusion along the edge competes with incorporation into the edge via mechanism (iii-a), whose barriers similarly increase from early to late TMs, ranging from 32~kJ~mol$^{-1}$ for Zr on Cu up to about 100~kJ~mol$^{-1}$ for Pd on Cu.
Although incorporation into the edge is therefore possible, particularly for early and central TMs, the barriers are considerable.
Crucially, however, for all but the latest TMs, the barrier for incorporation from below is still smaller than the energy required to detach from a step edge and even more so from a kink.
This means that once an adatom has attached, it can either remain at the edge, diffuse along it until it encounters a kink, or incorporate into the edge.

When an adatom approaches a step edge from the upper terrace, incorporation is generally much easier than incorporation from below.
Consistent with previous studies,\cite{wang_surface_2020, ali_mechanism_2022, kim_transition-pathway_2007-1, bellisario_importance_2009, trushin_step_1997} the mechanisms (ii-a), (iv-a), and (iv-b), in which the adatom replaces a host atom directly in the step edge or at a kink, are (nearly) barrierless for most dopant elements, except for the very latest TMs.
Once the adatom is sufficiently close to the defect, it can form bonds with the undercoordinated host atoms at the edge or kink, while fewer bonds of the displaced host atom must be broken.
This stabilises the transition state and enables essentially immediate incorporation.

Because adatoms perform a random walk across terraces, approach from above is just as likely as approach from below, and incorporation from above can therefore occur as an irreversible step, analogous to the irreversible attachment of adatoms to defects from the lower terrace.
Although the reactant state for incorporation from above is thermodynamically much less stable, the decisive factor is how the adatom arrives at the defect rather than the relative energies of the initial states.

An important exception to this general picture is Ag.
For Ag adatoms on Cu, the energy required for detachment from a step edge can be as small as 44~kJ~mol$^{-1}$, which is lower than the incorporation barrier of 83~kJ~mol$^{-1}$.
This implies that attachment of Ag to a step edge can be reversible, in contrast to the behaviour of most other 4\textit{d} TMs.
In addition, the barrier for hopping down a step edge is small, 42~kJ~mol$^{-1}$, comparable to the lowest barriers for incorporation from above via mechanism (iv-a).
Together, these properties indicate that Ag adatoms remain unusually mobile: they can detach from step edges, hop between terraces, and do not necessarily become trapped or promptly incorporated.
This enhanced mobility is consistent with experimental observations for Ag adatoms on Cu.\cite{bellisario_importance_2009}

Independent of the dominant pathways an adatom followed to become incorporated, it will be located in the upper terrace of the step edge, so continued incorporation leads to a locally higher dopant concentration.
As the step grows, the dopant build-up manifests as a dopant-rich brim, in agreement with experimental observations.\cite{bach_aaen_submonolayer_1998, lucci_atomic_2014, shi_formation_2021, bellisario_importance_2009, patel_elucidating_2019, patel2019atomic}
Thus, depending on the dopant elements, the diffusion and incorporation of subsequently deposited adatoms may be affected by the brim.

\begin{figure*}[ht]
\centering
  \includegraphics[width=\textwidth]{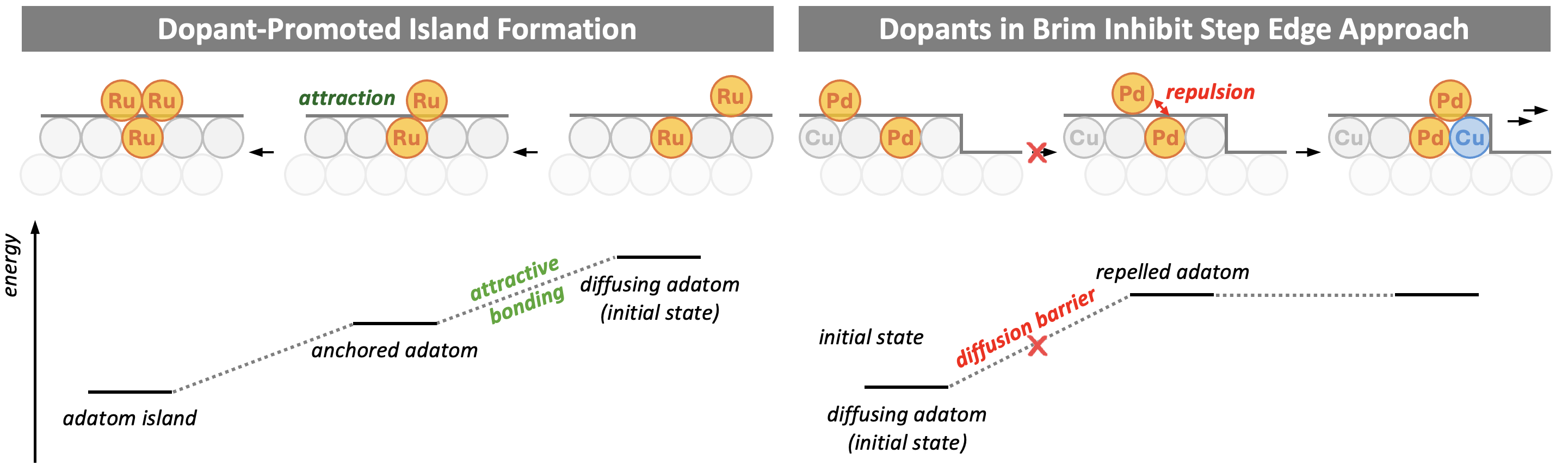}
  \caption{
    Schematic illustration of the effects of embedded dopant atoms on the diffusion and incorporation of adatoms.
    Dopant-promoted island formation: Attractive interactions between adatoms and dopants can immobilise adatoms on the surface, serving as anchor points for the formation of protruding adatom islands or facilitating the formation of embedded dopant clusters.
    Dopants in brim inhibit step edge approach: Repulsive interactions between adatoms and dopants can suppress incorporation mechanisms when a brim with a high local dopant concentration in the upper terrace next to a step edge prevents diffusing adatoms from approaching the edge from above.
    Schematic energy profiles are shown beneath the structural representations.
    The adatom and dopant are shown in orange, the moving host metal atoms in blue, and the adjacent host metal atoms in grey.
    Host atoms in the lowest terrace are shown in light grey, those in the central terrace in grey, and those in the uppermost terrace in dark grey.
    The surface, step edge, and kink are indicated with dark grey lines.
    Surface dimensions are for illustrative purposes only; the actual simulation cell sizes are provided in Section~S1 of the Supporting Information.
  }
  \label{fgr:dopant_presence}
\end{figure*}

\textbf{Dopant effects on adatom incorporation.}
The mechanistic picture developed so far does not account for how dopants that are already embedded in the surface influence the diffusion and incorporation of subsequently deposited adatoms.
Depending on the dopant element, embedded dopants can either attract or repel additional adatoms, and, as we will exemplify, these interactions can strongly affect whether adatoms incorporate, remain mobile, or form small adatom islands.
Figure~\ref{fgr:dopant_presence} illustrates these effects for both attractive and repulsive adatom–dopant interactions.

Very late TMs, such as Pd, do not form stabilising bonds with each other,\cite{berger_when_2025} so embedded dopants are expected to repel adatoms.
This has important consequences and explains surprising experimental observations for Pd in Cu.\cite{bach_aaen_submonolayer_1998,tierney_atomic-scale_2009,patel2019atomic}
Experiments show that the width of the Pd-rich brim in the upper terrace, and thus the amount of incorporated Pd, correlates with the area of the lower terrace connected to the step edge, but not with the area of the upper terrace.
This indicates that incorporation occurs exclusively via pathways in which adatoms approach the step from below, whereas incorporation from above is effectively suppressed.

This apparent contradiction to our mechanistic picture, which suggests that incorporation should occur from both the upper and lower terrace, can be rationalised by the repulsive interaction between Pd adatoms and embedded Pd dopants.
At low dopant concentrations, adatom diffusion across terraces is only weakly perturbed, as adatoms can diffuse around isolated dopants.
However, once a brim with a high local Pd concentration has formed, crossing this region becomes energetically unfavourable.
The diffusion barrier for crossing near a Pd dopant is sizeable, 33~kJ~mol$^{-1}$, compared with only 3~kJ~mol$^{-1}$ on pure Cu.
As a result, adatoms no longer perform a random walk but avoid Pd-rich regions, effectively creating an exclusion zone behind such areas.

Because this repulsion prevents adatoms from entering and crossing the dopant-rich upper-terrace brim, the reactant state required for low-barrier incorporation from above cannot form.
Consequently, incorporation proceeds exclusively via attachment from below, consistent with the experimental observation that the brim width correlates with the size of the lower, but not the upper, terrace connected to a step edge.

In contrast to the repulsive behaviour found for Pd, attractive adatom--dopant interactions can immobilise diffusing adatoms.
Ru is an example.
While diffusion of a Ru adatom across a clean terrace is very fast, with barriers below 4~kJ~mol$^{-1}$, its attachment to an embedded Ru dopant is strongly exothermic, releasing 116~kJ~mol$^{-1}$ on Cu and 198~kJ~mol$^{-1}$ on Ag.
This strong stabilisation fully immobilises the adatom, even when it binds from the terrace side above the dopant.
Such trapped adatoms may promote the formation of adatom islands on terraces, as illustrated in Figure~\ref{fgr:dopant_presence}.

Because these adatoms no longer diffuse, they cannot reach step edges or kink sites to incorporate via the dominant pathways.
We therefore investigate incorporation on a terrace next to the dopant via the concerted mechanism (i-b), which exhibits the lowest barriers on pure host terraces.
The barrier for incorporating a Ru adatom adjacent to an embedded Ru dopant is 91~kJ~mol$^{-1}$, substantially higher than on the pristine host surface, 59~kJ~mol$^{-1}$, and is therefore kinetically hindered.
This behaviour can be rationalised by the bonding changes in the transition state.
The Ru--Ru bond formed upon adsorption at the dopant remains intact, so no additional stabilisation is gained, while a dopant--host bond must be broken to accommodate the incoming adatom in the surface next to the existing dopant.
Since Ru forms stronger bonds with Cu than Cu does with itself, breaking a Ru--Cu bond incurs a larger energetic penalty than breaking a Cu--Cu bond on a pure host surface, leading to the increased incorporation barrier.
As a consequence, although Ru dopants strongly attract and immobilise Ru adatoms on the surface, the formation of embedded Ru dimers or larger dopant clusters may be kinetically self-inhibiting.
Reaction energies for adatom attachment to embedded dopants and diffusion barriers over embedded dopants are reported in Section~S3 in the Supporting Information.

\section{Conclusion}

In summary, this work provides a broad and coherent mechanistic understanding of how dopant adatoms evolve into embedded active sites in single-atom alloys.
Using DFT, we identify and rationalise periodic trends across the 4\textit{d} TM series and establish the dominant incorporation pathways.
Direct incorporation into terrace sites is generally unfavourable, although a concerted three-atom mechanism may offer a viable route at elevated temperatures or on sufficiently extended terraces.

Surface defects, in contrast, enable efficient adatom incorporation.
Both step edges and kink sites facilitate (almost) barrierless incorporation for most TMs when approached from above, with step edges alone already providing low-barrier pathways.
Attachment to step edges, and particularly to kink sites, from the lower terrace is thermodynamically strongly favoured and effectively irreversible for most TMs.
Once attached, adatoms may initiate step growth or incorporate into the edge, as incorporation barriers are usually smaller than those for detachment.
Ag represents an exception.
On Cu surfaces, detachment of Ag adatoms from step edges is easier than incorporation, and the combination of comparatively high incorporation barriers and low hop-down barriers makes Ag adatoms unusually mobile, allowing them to move between terraces.

Periodic trends reveal that early and central TMs incorporate most readily, whereas very late TMs exhibit higher barriers for the relevant pathways.
Overall, Cu hosts display slightly higher incorporation barriers and less exothermic reaction energies than Ag, reflecting stronger host--host bonding and a larger size mismatch with 4\textit{d} dopants.

Finally, adatom--dopant interactions can significantly modulate diffusion and incorporation.
Repulsive interactions, as in Pd-doped Cu, suppress incorporation from above by creating dopant-rich brims that act as exclusion zones, preventing adatoms from reaching the edge to form the reactant states needed.
Attractive interactions, exemplified by Ru, immobilise diffusing adatoms at embedded dopants and can promote the formation of anchored adatom islands, yet the formation of embedded dopant dimers or larger clusters is kinetically inhibited.

Together, these findings bridge the gap between elementary incorporation mechanisms and experimentally observed synthesis outcomes, providing mechanistic insight into SAA formation and guidance on which surface environments most effectively promote incorporation for different dopant elements, as well as when specific dopant--adatom interactions suppress particular incorporation pathways.

\section*{Conflicts of interest}
There are no conflicts to declare.

\section*{Data availability}
The data supporting the findings of this study 
are provided in the article and in the Supporting Information.

\section*{Acknowledgements}
I.K. acknowledges support from the Alexander S. Onassis Foundation through an Onassis Foundation Scholarship, from Christ’s College, Cambridge, through a College Grant, and from the Royal Society of Chemistry through an Undergraduate Research Bursary (U25-9082313565).
F.B. acknowledges support from the Alexander von Humboldt Foundation through a Feodor Lynen Research Fellowship, from the Isaac Newton Trust through an Early Career Fellowship, and from Churchill College, Cambridge, through a Postdoctoral By-Fellowship. 
This work has been funded by the European Union (ERC, n-AQUA, 101071937). 
Views and opinions expressed are, however, those of the authors only and do not necessarily reflect those of the European Union or the European Research Council Executive Agency. 
Neither the European Union nor the granting authority can be held responsible for them. 
This work was performed using resources provided by the Cambridge Service for Data Driven Discovery (CSD3) operated by the University of Cambridge Research Computing Service (www.csd3.cam.ac.uk), provided by Dell EMC and Intel using Tier-2 funding from the Engineering and Physical Sciences Research Council (capital grant EP/T022159/1), and DiRAC funding from the Science and Technology Facilities Council (www.dirac.ac.uk), with additional access through a University of Cambridge EPSRC Core Equipment Award (EP/X034712/1). 
We additionally acknowledge computational support and resources from the UK National High-Performance Computing Service, Advanced Research Computing High End Resource (ARCHER2). 
Access for ARCHER2 was obtained via the Materials Chemistry Consortium (MCC), funded by EPSRC grant references EP/X035859 and EP/F067496. Further computational support and resources were provided by YOUNG, the Tier-2 High Performance Computing Hub in Materials and Molecular Modeling (MMM), which is partially funded by EPSRC grant reference EP/T022213.
We also acknowledge EuroHPC Joint Undertaking for awarding the project ID EHPC-REG-2024R02-130 access to Leonardo at CINECA, Italy.

\clearpage

\bibliography{refs}

\end{document}


\clearpage

\tableofcontents

\clearpage

\section{Computational Details}
Periodic density functional theory (DFT) calculations with the optB86b-vdW functional\cite{optB86b-vdW} are performed as implemented in VASP 6.4.1, using plane wave basis sets for valence electrons and the projector augmented wave (PAW) method for core electrons with standard potentials.\cite{kresse1993ab,kresse1996efficiency,kresse1996efficient,kresse1999ultrasoft} 
Further, first order Methfessel-Paxton smearing (SIGMA = 0.1), dipole correction, and the following POTCARs are used: 
Cu: Cu, Ag: Ag, Y: Y\textunderscore sv, Zr: Zr\textunderscore sv, Nb: Nb\textunderscore sv, Mo: Mo\textunderscore sv, Tc: Tc\textunderscore pv, Ru: Ru\textunderscore pv, Rh: Rh\textunderscore pv, Pd: Pd.

The metal surfaces are modelled using five atomic layers, with the bottom two layers fixed, and a vacuum region of 15~\AA.
Simulation cell sizes are chosen such that dopants or adatoms are separated by at least two host atoms.
For the (111) facet, calculations are carried out in (3~$\times$~3~$\times$~5) and (4~$\times$~4~$\times$~5) supercells.
For the (211) facet, cells containing three atom rows along each lateral direction are used.
For the (322) facet, we employ cells with three and five atom rows along the lateral directions.
Kink-site models are generated by attaching additional host atoms to the (322) step edge, using simulation cells with four and five atom rows along the lateral directions.
All structure files are provided as part of the Supporting Information.
 
Unit cell vectors are determined using the equation of state method\cite{alchagirov2003reply} as implemented in the Atomic Simulation Environment (ASE), an energy cutoff of 800 eV is used to minimise the influence of volume change on the basis set, and the Brillouin zone is sampled using a Monkhorst-Pack mesh with 31 $\times$ 31 $\times$ 31 \textbf{k} points.

For structure optimisations, the plane‐wave energy cutoff is set to 400~eV, and the Brillouin zone is sampled using a 5~$\times$~5~$\times$~1 Monkhorst–Pack \textbf{k}-point mesh.\cite{monkhorstpack}
Structures are relaxed using the conjugate‐gradient algorithm until all forces are below 0.01~eV/\AA.

Transition state structures are first located using the Nudged Elastic Band (NEB) method as implemented in the VTST tools for VASP,\cite{nebimproved} employing 9 images with the initial and final images fixed.
NEB optimisations are stopped when all atomic forces are below 0.1~eV/\AA.
The images with the highest are then interpolated to construct initial guesses for the dimer method, which is subsequently used to refine the transition state structure.\cite{dimer}
Dimer calculations are converged when all forces are below 0.01~eV/\AA.

\clearpage

\section{Periodic Trends in Reaction Energies and Barriers}

\begin{table}
  \centering
  \renewcommand{\arraystretch}{1.5} 
  \caption{
    \textbf{Reaction energies} for mechanisms (i), (ii), (iii), (iv), and (v), comprising adatom \textbf{diffusion} on terraces (fcc to hcp), diffusion along step edges, hopping down a descending step edge, and \textbf{attachment} to step edges and kink sites on \textbf{Cu surfaces}. 
    Energies are reported in kJ mol$^{-1}$ for the 3d TM Cu and the series of 4d TM adatoms. 
    Reaction energies are defined as the energy change from the initial to the final state as described in the main text.
  }
  \vspace{8pt}
  \begin{tabular}{@{\hspace{8pt}}*{13}{c@{\hspace{8pt}}}}
    \hline\hline
    mechanism & $|$ & Cu & $|$ & Y & Zr & Nb & Mo & Tc & Ru & Rh & Pd & Ag \\
    \hline
    (i) &  & $ 1$ & & $0$ & $-2$ & $-2$ & $-1$ & $2$ & $3$ & $4$ & $2$ & $0$ \\
    (ii) &  & $-60$ &  & $ $ & $-118$ & $-123$ &    & $-112$ & $-106$ & $-95$ & $-78$ & $-50$ \\
    (iii) &  & $-55$ &  & $ $ & $-99$ & $-111$ & $-115$ & $-115$ & $-112$ & $-104$ & $-81$ & $-44$ \\
    (iv) &  & $0$ &  & $0$ & $0$ & $0$ & $0$ & $0$ & $0$ & $0$ & $0$ & $0$ \\
    (v) &  & $-79$ &  & $ $ & $ $ & $-170$ & $-172$ & $-171$ & $-161$ & $-146$ & $-112$ & $-63$ \\ 
    \hline\hline
  \end{tabular}
  \label{tab:example}
\end{table}

\begin{table}
  \centering
  \renewcommand{\arraystretch}{1.5} 
  \caption{
    \textbf{Reaction barriers} for mechanisms (i), (ii), and (iv), comprising adatom \textbf{diffusion} on terraces (fcc to hcp), diffusion along step edges, and hopping down a descending step edge on \textbf{Cu surfaces}. 
    Barriers for attachment to step edges and kink sites are not reported, as they are lower than the terrace diffusion barriers. 
    Energies are reported in kJ~mol$^{-1}$ for the 3d TM Cu and the series of 4d TM adatoms. 
    Reaction barriers are defined as the energy change from the initial to the transition state as described in the main text.
  }
  \vspace{8pt}
  \begin{tabular}{@{\hspace{8pt}}*{13}{c@{\hspace{8pt}}}}
    \hline\hline
    mechanism & $|$ & Cu & $|$ & Y & Zr & Nb & Mo & Tc & Ru & Rh & Pd & Ag \\
    \hline
    (i) &  & $5$ & & $1$ & $2$ & $4$ & $4$ & $5$ & $3$ & $4$ & $3$ & $4$ \\
    (ii) &  & $49$ & & $ $ & $104$ & $126$ &  & $134$ & $123$ & $108$ & $77$ & $42$ \\
    (iii) &  & - & & - & - & - & - & - & - & - & - & - \\
    (iv) &  & $30$ & & $12$ & $21$ & $26$ & $33$ & $36$ & $41$ & $42$ & $35$ & $19$ \\
    (v) &  & - & & - & - & - & - & - & - & - & - & - \\
    \hline\hline
  \end{tabular}
  \label{tab:example}
\end{table}

\begin{table}
  \centering
  \renewcommand{\arraystretch}{1.5} 
  \caption{
    \textbf{Reaction energies} for mechanisms (i), (ii), (iii), (iv), and (v), comprising adatom \textbf{diffusion} on terraces (fcc to hcp), diffusion along step edges, hopping down a descending step edge, and \textbf{attachment} to step edges and kink sites on \textbf{Ag surfaces}. 
    Energies are reported in kJ mol$^{-1}$ for the 3d TM Cu and the series of 4d TM adatoms. 
    Reaction energies are defined as the energy change from the initial to the final state as described in the main text.
  }
  \vspace{8pt}
  \begin{tabular}{@{\hspace{8pt}}*{13}{c@{\hspace{8pt}}}}
    \hline\hline
    mechanism & $|$ & Cu & $|$ & Y & Zr & Nb & Mo & Tc & Ru & Rh & Pd & Ag \\
    \hline
    (i) &  & $1$ &  & $0$ & $-1$ & $-1$ & $1$ & $2$ & $3$ & $4$ & $3$ & $0$ \\
    (ii) &  & $-58$ &  & $ $ & $-109$ & $-113$ & $-111$ & $-108$ & $-103$ & $-94$ & $-84$ & $-52$ \\
    (iii) &  & $-54$ &  & $ $ & $-100$ & $-107$ & $-107$ & $-107$ & $-110$ & $-102$ & $-81$ & $-48$ \\
    (iv) &  & $0$ &  & $0$ & $0$ & $0$ & $0$ & $0$ & $0$ & $0$ & $0$ & $0$ \\
    (v) &  & $ $ &  & $ $ & $-147$ & $-158$ & $-160$ & $-157$ & $-151$ & $-138$ & $-110$ & $-69$ \\
    \hline\hline
  \end{tabular}
  \label{tab:example}
\end{table}

\begin{table}
  \centering
  \renewcommand{\arraystretch}{1.5} 
  \caption{
    \textbf{Reaction barriers} for mechanisms (i), (ii), and (iv), comprising adatom \textbf{diffusion} on terraces (fcc to hcp), diffusion along step edges, and hopping down a descending step edge on \textbf{Ag surfaces}. 
    Barriers for attachment to step edges and kink sites are not reported, as they are lower than the terrace diffusion barriers. 
    Energies are reported in kJ~mol$^{-1}$ for the 3d TM Cu and the series of 4d TM adatoms. 
    Reaction barriers are defined as the energy change from the initial to the transition state as described in the main text.
  }
  \vspace{8pt}
  \begin{tabular}{@{\hspace{8pt}}*{13}{c@{\hspace{8pt}}}}
    \hline\hline
    mechanism & $|$ & Cu & $|$ & Y & Zr & Nb & Mo & Tc & Ru & Rh & Pd & Ag \\
    \hline
    (i) &  & $7$ &  & $1$ & $2$ & $3$ & $4$ & $5$ & $3$ & $4$ & $5$ & $5$ \\
    (ii) &  & $42$ &  & $ $ & $98$ & $115$ & $120$ & $118$ & $110$ & $96$ & $65$ & $38$ \\
    (iii) &  & - & & - & - & - & - & - & - & - & - & - \\
    (iv) &  & $33$ &  & $22$ & $33$ & $38$ & $41$ & $44$ & $47$ & $49$ & $41$ & $24$ \\
    (v) &  & - & & - & - & - & - & - & - & - & - & - \\
    \hline\hline
  \end{tabular}
  \label{tab:example}
\end{table}

\begin{table}
  \centering
  \renewcommand{\arraystretch}{1.5} 
  \caption{
    \textbf{Reaction energies for incorporation mechanisms} (i-a), (i-b), (ii-a), (ii-b), (iii-a), (iii-b), (iv-a), and (iv-b), comprising adatom incorporation into terrace sites, step edges, and kink sites on \textbf{Cu surfaces}.
    Energies are reported in kJ~mol$^{-1}$ for the 3d TM Cu and the series of 4d TM adatoms.
    Reaction energies are defined as the energy change from the initial to the final state, as described in the main text.
    For mechanism (iii-b), a high-lying intermediate may exist; when this occurs, the reaction energies are reported separately for the step from the initial state to the intermediate (iii-b-1) and from the intermediate to the final product (iii-b-2).
    }
  \vspace{8pt}
  \begin{tabular}{@{\hspace{8pt}}*{13}{c@{\hspace{8pt}}}}
    \hline\hline
    mechanism & $|$ & Cu & $|$ & Y & Zr & Nb & Mo & Tc & Ru & Rh & Pd & Ag \\
    \hline
    (i-a) &  & $0$ & & $-24$ & $-109$ & $-161$ & $-184$ & $-190$ & $-169$ & $-125$ & $-47$ & $31$ \\
    (i-b) &  & $0$ & & $-30$ & $-119$ & $-145$ & $-163$ & $-172$ & $-150$ & $-116$ & $-51$ & $34$ \\
    (ii-a) &  & $-57$ & & $ $ & $-148$ & $-177$ & - & - & - & $-149$ & $-100$ & $-30$ \\
    (ii-b) &  & $-55$ & & $ $ & $-136$ & $-189$ & $-214$ & $-220$ & $-207$ & $-172$ & $-104$ & $-18$ \\
    (iii-a) &  & $58$ & & $36$ & $17$ & $0$ & $-8$ & $-12$ & $-4$ & $17$ & $50$ & $66$ \\
    (iii-b) &  & $0$ & & $5$ & $-33$ & $-55$ & $-68$ & $-75$ & $-69$ & $-52$ & $-19$ & $17$ \\
    (iii-b-1) &  & $75$ & & $77$ & $74$ & - & - & $73$ & $74$ & $73$ & $76$ & $74$ \\
    (iii-b-2) &  & $-75$ & & $-72$ & $-107$ & - & - & $-148$ & $-142$ & $-126$ & $-95$ & $-57$ \\
    (iv-a) &  & $-57$ & & $ $ & $-139$ & $-180$ & $-197$ & $-198$ & $-185$ & $-150$ & $-99$ & $-30$ \\
    (iv-b) &  & $-79$ & & $ $ & $ $ & $-208$ & $-216$ & $-218$ & $-204$ & $-176$ & $-122$ & $-52$ \\
    \hline\hline
  \end{tabular}
  \label{tab:example}
\end{table}

\begin{table}
  \centering
  \renewcommand{\arraystretch}{1.5} 
  \caption{
    \textbf{Reaction barriers for incorporation mechanisms} (i-a), (i-b), (ii-a), (ii-b), (iii-a), (iii-b), (iv-a), and (iv-b), comprising adatom incorporation into terrace sites, step edges, and kink sites on \textbf{Cu surfaces}.
    Energies are reported in kJ~mol$^{-1}$ for the 3d TM Cu and the series of 4d TM adatoms.
    Reaction barriers are defined as the energy change from the initial to the transition state, as described in the main text.
    For mechanism (iii-b), a high-lying intermediate may exist; when this occurs, the reaction barriers are reported separately for the step from the initial state to the intermediate (iii-b-1) and from the intermediate to the final product (iii-b-2).
    }
  \vspace{8pt}
  \begin{tabular}{@{\hspace{8pt}}*{13}{c@{\hspace{8pt}}}}
    \hline\hline
    mechanism & $|$ & Cu & $|$ & Y & Zr & Nb & Mo & Tc & Ru & Rh & Pd & Ag \\
    \hline
    (i-a) &  & $136$ & & $121$ & $99$ & $96$ & $96$ & $95$ & $98$ & $108$ & $131$ & $149$ \\
    (i-b) &  & $124$ & & $79$ & $51$ & $41$ & $42$ & $48$ & $59$ & $80$ & $103$ & $130$ \\
    (ii-a) &  & $38$ & & $ $ & $4$ & $9$ & $0$ & $0$ & $0$ & $9$ & $21$ & $46$ \\
    (ii-b) &  & $86$ & & $ $ & $46$ & $32$ & $27$ & $28$ & $33$ & $44$ & $66$ & $95$ \\
    (iii-a) &  & $97$ & & $43$ & $32$ & $38$ & $47$ & $54$ & $64$ & $83$ & $101$ & $ $ \\
    (iii-b) &  & $81$ & & $86$ & $79$ & $76$ & $75$ & $74$ & $77$ & $82$ & $86$ & $83$ \\
    (iii-b-1) &  & $81$ & & $86$ & $79$ & - & - & $74$ & $74$ & $73$ & $77$ & $82$ \\
    (iii-b-2) &  & $6$ & & $1$ & $0$ & - & - & $0$ & $4$ & $9$ & $10$ & $8$ \\
    (iv-a) &  & $30$ & & $ $ & $ $ & $11$ & $10$ & $10$ & $8$ & $10$ & $14$ & $42$ \\
    (iv-b) &  & $38$ & & $ $ & $ $ & $0$ & $0$ & $8$ & $2$ & $7$ & $22$ & $45$ \\
    \hline\hline
  \end{tabular}
  \label{tab:example}
\end{table}

\begin{table}
  \centering
  \renewcommand{\arraystretch}{1.5} 
  \caption{
    \textbf{Reaction energies for incorporation mechanisms} (i-a), (i-b), (ii-a), (ii-b), (iii-a), (iii-b), (iv-a), and (iv-b), comprising adatom incorporation into terrace sites, step edges, and kink sites on \textbf{Ag surfaces}.
    Energies are reported in kJ~mol$^{-1}$ for the 3d TM Cu and the series of 4d TM adatoms.
    Reaction energies are defined as the energy change from the initial to the final state, as described in the main text.
    For mechanism (iii-b), a high-lying intermediate may exist; when this occurs, the reaction energies are reported separately for the step from the initial state to the intermediate (iii-b-1) and from the intermediate to the final product (iii-b-2).
  }
  \vspace{8pt}
  \begin{tabular}{@{\hspace{8pt}}*{13}{c@{\hspace{8pt}}}}
    \hline\hline
    mechanism & $|$ & Cu & $|$ & Y & Zr & Nb & Mo & Tc & Ru & Rh & Pd & Ag \\
    \hline
    (i-a) &  & $-18$ & & $-91$ & $-169$ & $-201$ & $-206$ & $-202$ & $-183$ & $-147$ & $-81$ & $0$ \\
    (i-b) &  & $-3$ & & $-76$ & $-146$ & $-176$ & $-179$ & $-173$ & $-158$ & $-126$ & $-72$ & $0$ \\
    (ii-a) &  & $-55$ & & $-114$ & $-171$ & $-193$ & $-195$ & $-190$ & $-176$ & $-151$ & $-111$ & $-50$ \\
    (ii-b) &  & $-54$ & & $-114$ & $-189$ & $-218$ & $-223$ & $-219$ & $-210$ & $-177$ & $-119$ & $-48$ \\
    (iii-a) &  & $39$ & & $11$ & $-13$ & $-28$ & $-32$ & $-31$ & $-21$ & $-2$ & $26$ & $49$ \\
    (iii-b) &  & $0$ & & $-31$ & $-58$ & $-75$ & $-78$ & $-78$ & $-71$ & $-57$ & $-29$ & $0$ \\
    (iii-b-1) &  & $59$ & & $46$ & $55$ & - & - & $59$ & $63$ & $62$ & $63$ & $58$ \\
    (iii-b-2) &  & $-59$ & & $-77$ & $-113$ & - & - & $-137$ & $-134$ & $-119$ & $-92$ & $-58$ \\
    (iv-a) &  & $ $ & & $ $ & $-164$ & $-194$ & $-194$ & $-200$ & $-181$ & $-152$ & $-107$ & $-46$ \\
    (iv-b) &  & $ $ & & $ $ & $-189$ & $-209$ & $-212$ & $-207$ & $-194$ & $-170$ & $-126$ & $-69$ \\
    \hline\hline
  \end{tabular}
  \label{tab:example}
\end{table}

\begin{table}
  \centering
  \renewcommand{\arraystretch}{1.5} 
  \caption{
    \textbf{Reaction barriers for incorporation mechanisms} (i-a), (i-b), (ii-a), (ii-b), (iii-a), (iii-b), (iv-a), and (iv-b), comprising adatom incorporation into terrace sites, step edges, and kink sites on \textbf{Ag surfaces}.
    Energies are reported in kJ~mol$^{-1}$ for the 3d TM Cu and the series of 4d TM adatoms.
    Reaction barriers are defined as the energy change from the initial to the transition state, as described in the main text.
    For mechanism (iii-b), a high-lying intermediate may exist; when this occurs, the reaction barriers are reported separately for the step from the initial state to the intermediate (iii-b-1) and from the intermediate to the final product (iii-b-2).
  }
  \vspace{8pt}
  \begin{tabular}{@{\hspace{8pt}}*{13}{c@{\hspace{8pt}}}}
    \hline\hline
    mechanism & $|$ & Cu & $|$ & Y & Zr & Nb & Mo & Tc & Ru & Rh & Pd & Ag \\
    \hline
    (i-a) &  & $100$ & & $78$ & $72$ & $77$ & $80$ & $80$ & $83$ & $90$ & $107$ & $112$ \\
    (i-b) &  & $103$ & & $53$ & $38$ & $36$ & $37$ & $42$ & $51$ & $65$ & $86$ & $108$ \\
    (ii-a) &  & $28$ & & $16$ & $1$ & $1$ & $0$ & $1$ & $3$ & $10$ & $16$ & $35$ \\
    (ii-b) &  & $68$ & & $48$ & $28$ & $24$ & $24$ & $25$ & $25$ & $35$ & $54$ & $76$ \\
    (iii-a) &  & $76$ & & $31$ & $36$ & $42$ & $48$ & $53$ & $64$ & $76$ & $87$ & $86$ \\
    (iii-b) &  & $66$ & & $66$ & $64$ & $65$ & $65$ & $65$ & $65$ & $69$ & $73$ & $67$ \\
    (iii-b-1) &  & $66$ & & $66$ & $64$ & - & - & $65$ & $65$ & $64$ & $67$ & $67$ \\
    (iii-b-2) &  & $5$ & & $11$ & $1$ & - & - & $0$ & $1$ & $6$ & $10$ & $9$ \\
    (iv-a) &  & $ $ & & $ $ & $0$ & $0$ & $0$ & $0$ & $0$ & $4$ & $10$ & $31$ \\
    (iv-b) &  & $ $ & & $ $ & $ $ & $0$ & $0$ & $0$ & $2$ & $7$ & $18$ & $35$ \\
    \hline\hline
  \end{tabular}
  \label{tab:example}
\end{table}

\clearpage

\section{Reaction Energies and Barriers for Dopant--Adatom Interactions}

The reaction energy for the diffusion of a Pd adatom from an fcc to an hcp site when crossing an embedded Pd dopant in a Cu surface is 2~kJ~mol$^{-1}$, and the corresponding barrier is 33~kJ~mol$^{-1}$.
On pure Cu, the reaction energy is the same, 2~kJ~mol$^{-1}$, but the barrier is much lower at 3~kJ~mol$^{-1}$, indicating that Pd adatoms will avoid diffusing across embedded Pd atoms.

The energy released when a Ru adatom attaches from a pure host terrace site onto an embedded Ru dopant is $-198$~kJ~mol$^{-1}$ on Cu and $-116$~kJ~mol$^{-1}$ on Ag.
The reaction energy for incorporating a Ru adatom next to an embedded Ru dopant in Cu is $-69$~kJ~mol$^{-1}$, which is substantially less exothermic than the incorporation of the first Ru adatom into a terrace site, $-169$~kJ~mol$^{-1}$.
The barrier for incorporating a Ru adatom adjacent to an embedded Ru dopant in Cu is 91~kJ~mol$^{-1}$, 32~kJ~mol$^{-1}$ higher than the barrier for incorporation of the first Ru atom on a pure Cu terrace, which is 59~kJ~mol$^{-1}$.
This shows that although incorporating a second Ru atom is thermodynamically favourable, the process is associated with a high barrier; even higher than the incorporation barrier on the pristine surface, indicating that the formation of embedded dopant clusters may be kinetically hindered.

Finally, the attachment of a second Ru adatom to a Ru adatom already anchored at an embedded Ru dopant, as illustrated in Figure~5 of the main text, is also highly exothermic, releasing $-152$~kJ~mol$^{-1}$.
This demonstrates that embedded dopants can trap additional adatoms when the element exhibits attractive self-interactions, potentially promoting the formation of adatom islands on terraces.

\clearpage
\bibliography{refs}